\documentclass{aa}   
\usepackage{graphicx,epsfig,txfonts}   
\begin{document}   
   
\newcommand{\ea}{{et al.}}   
\newcommand{\beq}{\begin{equation}}   
\newcommand{\enq}{\end{equation}}   
\newcommand{\bfg}{\begin{figure}}   
\newcommand{\efg}{\end{figure}}   
\newcommand{\bfa}{\begin{figure*}}   
\newcommand{\efa}{\end{figure*}}   
\newcommand{\bea}{\begin{eqnarray}}   
\newcommand{\ena}{\end{eqnarray}}   
\newcommand{\dd}{{\rm{d}}}   
\newcommand{\dg}{^{\rm{o}}}   
\newcommand{\thetao}{{\theta_{\rm{o}}}}   
\newcommand{\nuo}{{\nu_{\rm{o}}}}   
\newcommand{\thetae}{{\theta_{\rm{e}}}}   
\newcommand{\nue}{{\nu_{\rm{e}}}}   
\newcommand{\re}{{r_{\rm{e}}}}   
\newcommand{\rin}{{r_{\rm{in}}}}   
\newcommand{\rout}{{r_{\rm{out}}}}   
\newcommand{\const}{{\mbox{const}}}   
\newcommand{\bmath}[1]{\mbox{\boldmath{${#1}$}}}   
   
\def\lb#1{{\protect\linebreak[#1]}}   
   
\def\apj{ApJ}   
\def\apjs{ApJS}   
\def\asa{A\&A}   
\def\mnras{MNRAS}   
\def\phr{Phys. Rev.}   
\def\phrl{Phys. Rev. Lett.}   
   
\def\ltsima{$\; \buildrel < \over \sim \;$}   
\def\gtsima{$\; \buildrel > \over \sim \;$}   
\def\simlt{\lower.5ex\hbox{\ltsima}}   
\def\simgt{\lower.5ex\hbox{\gtsima}}   
\def\ebf{{\bf e}}   
\def\Ebf{{\bf E}}   
\def\dpar{{\bf \partial}}   
\def\ka{{\rm K}$\alpha$}   
\def\kb{{\rm K}$\beta$}   
\def\ee{\`{e}}   
   
\title{The XMM-{\fontshape{it}\selectfont{}Newton} view of GRS\,1915+105\thanks{Based
on observations obtained with XMM-{\fontshape{it}\selectfont{}Newton}, an ESA
science mission with instruments and contributions directly funded by ESA Member
States and NASA. }}   
\author{
A.\ Martocchia\inst{1}\thanks{{\it Present address:} UPS / Centre d'Etude Spatiale des Rayonnements,
9 Av. du Colonel Roche, F--31028 Toulouse, France}, 
G.\ Matt\inst{2}, 
T.\ Belloni\inst{3}, 
M.\ Feroci\inst{4},    
V.\ Karas\inst{5}, 
G.\ Ponti\inst{6,7,8}
}
\offprints{A.\ Martocchia, {\sf andrea.martocchia@cesr.fr}}       
\institute{   
CNRS / Observatoire Astronomique de Strasbourg, 11 Rue de l'Universit\'e,    
F--67000 Strasbourg, France   
\and   
Dipartimento di Fisica, Universit\`a degli Studi ``Roma Tre'',   
Via della Vasca Navale 84, I--00146 Roma, Italy   
\and   
INAF / Osservatorio Astronomico di Brera, via E. Bianchi 46, I--23807 Merate, Italy   
\and   
INAF / IASF, Area di Ricerca di Tor Vergata, Via Fosso del Cavaliere 100,    
I--00133 Roma, Italy    
\and   
Astronomical Institute, Academy of Sciences,    
Bo\v{c}n\'{\i}~II, CZ--140\,31~Prague, Czech Republic   
\and   
INAF / IASF, Sezione di Bologna, via Gobetti 101, I--40129 Bologna, Italy   
\and     
Dipartimento di Astronomia, Universit\`a di Bologna, via Ranzani 1,    
I--40127 Bologna, Italy   
\and    
Institute of Astronomy, Madingley Road, Cambridge CB3 0HA, United Kingdom    
}      
\date{Received... Accepted...}

\abstract{Two XMM-{\it Newton} observations of the black-hole binary   
GRS\,1915+105 were triggered in 2004 (April 17 and 21),   
during a long ``plateau'' state of the source.    
We analyzed the data collected with EPIC-pn in timing and burst modes,   
respectively. Reflection Grating Spectrometers were used only on April 21st.    
The source 2--10 keV flux is $\sim 0.6$ (unabsorbed: $0.9\div1.1$) $\times    
10^{-8}$ in cgs units.   
While the light curves show only small amplitude variations (a few percent)   
at timescales longer than a few seconds, a QPO is seen at about   
0.6 Hz -- as expected in $\chi$ variability modes of   
GRS\,1915+105, when the phenomenological correlation   
with the source flux is taken into account -- possibly with a harmonic    
signal at 1.2 Hz. The pn spectrum is well fitted without
invoking thermal disk emission, on the basis of four main components:    
a primary one (either a simple power law or thermal Comptonization models), 
absorbed by cold matter with abundances different than those of standard ISM;    
reprocessing from an ionized disk; emission and absorption lines; and 
a soft X-ray excess at $\sim 1$ keV.  
However, the last is not confirmed by the RGS spectra, whose difference
from the EPIC-pn ones lacks a fully satisfactory explanation. 
If real, the soft X-ray excess may be due 
to reflection from an optically thin, photoionized disk wind; in this case
it may lead to a way to disentangle intrinsic from interstellar absorption. 
 }   
   
\authorrunning{A. Martocchia et al.}   
\titlerunning{The XMM-{\it Newton} view of GRS\,1915+105 }   
\maketitle   
   
        \section{Introduction}   
        \label{sec:intro}   
   
The well-known    
black-hole (BH) binary GRS\,1915+105 is sometimes classified as a 
{\it superluminal microquasar}    
because of the impressive radio jets whose matter seems to move with   
superluminal velocity due to a relativistic effect (Mirabel \& Rodriguez, 1994;   
Rodriguez \& Mirabel, 1999; Fender \ea, 1999).    
For a recent review of this source, including a general discussion of its    
properties, see Fender \& Belloni (2004).   
   
Due to very large obscuration, the spectral type of GRS\,1915+105's    
companion (a K-M III star) was discovered only recently, via    
infrared observations (Greiner \ea, 2001). This yielded evidence    
that the system, initially discovered in the $\gamma$ rays     
(Castro-Tirado \ea\ 1992), belongs to the class of low-mass X-ray binaries    
(LMXBs). On the basis of the commonly adopted    
value of the source's inclination ($70^\circ$; but see Maccarone, 2002,    
for a discussion),  Greiner et al. (2001b) used the same observations to determine the    
mass of the central compact object, which has been constrained to   
$M_{\rm c}=14\pm4 M_\odot$, i.e. well above the maximum   
neutron star mass limit.  Thus  
GRS\,1915+105 is believed to host a BH with a gravitational    
radius $r_{\rm g}=\frac{GM}{c^2} \sim 21$ km, so it is    
classified as a BH binary.   
  
Reig \ea\ (2003) note that,   
although the phenomenology of GRS\,1915+105 has been   
usually\footnote{Based on the hardness ratio and position in the   
color-color diagram. } described in terms of three spectral states   
named A, B, and C (the latter being low/hard and jet-dominated), in fact several    
source properties always correspond to the canonical {\it intermediate}   
(sometimes called {\it very high}) {\it state} of galactic BH sources   
(see Homan \& Belloni, 2005, and Belloni et al., 2005, for a definition).   
This is confirmed by Done \ea\ (2004),    
who also note that the source's unique limit-cycle variability   
appears when the source radiates at super-Eddington luminosities.   
Many of the GRS\,1915+105 properties can be related to the   
binary's evolutionary state: the giant companion star and the long   
orbital period (33.4 days, by far the longest of any LMXB    
(Greiner \ea, 2001)   
indicate that the Roche lobe overflow occurs in a very wide binary.   
The cause of all the unique long, as well as short-term, variability   
of GRS\,1915+105 is then   
linked to the evolution of the huge disk structure, which can contain   
enough material to mantain nearly-Eddington accretion rates over timescales   
of several years.   
   
A strong and broad iron \ka\ line was present in {\it Beppo}SAX  
(Martocchia \ea\ 2002, 2004) and {\it ASCA} GIS (Miller \ea\ 2004)   
spectra of GRS\,1915+105.   
{\it Rossi}-XTE PCA data are generally   
well-fitted including a broad iron line component (e.g. Done et al. 2004),    
but the energy resolution of this instrument does not allow discriminating    
among different line models. The  
Fe K$\alpha$ emission was also compatible with the spectral fits of   
{\it Chandra} data, taken during   
a low/hard state of the source, as reported by Lee \ea\ (2002). 

An XMM-{\it Newton} ToO observation of GRS\,1915+105, which would   
allow us to study the iron line profile with enhanced spectral    
resolution and sensitivity, was proposed in the second Announcement of 
Opportunity (AO2). The observation was intended to be triggered    
by the occurrence of a ``plateau'' state of the source (Foster \ea\ 1996),    
similar to the one observed during the     
{\it Beppo}SAX 1998 observation; this was also necessary in order to have   
the source in a less dramatic variability state and at a lower   
flux level to minimize technical problems due to instrumental pile-up   
and telemetry.    
For the latter reason we also planned to observe in timing mode.    
   
The observation was triggered in April 2004, divided into   
two parts with $\sim 4$ days separation one from the other.   
In this paper, while presenting the results,   
we offer an XMM-{\it Newton} view of GRS\,1915+105 for the first time.   
First of all we describe (in Sect.~\ref{sec:reduction}) the   
observations and data extraction methods, and details of the state of the    
source based on its variability and general considerations.   
In Sect.~\ref{sec:spectral} the best-fit spectral models and    
related physical parameters are given. Finally, in the   
conclusions, we summarize these XMM-{\it Newton} measurements    
and put our results in the context of the known    
source properties and open issues.   
   
\begin{figure}   
\centering   
\hfill   
\includegraphics[angle=-90,width=0.48\textwidth]{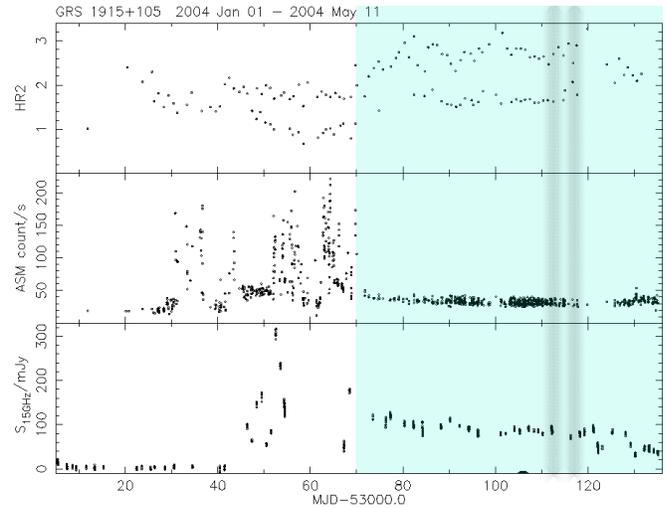}   
\caption{ Radio and X-ray light curves of GRS\,1915+105 before and   
during the Spring 2004 plateau (starting at $\sim$ 2453070 JD),   
from: http://\lb{2}www.mrao.cam.ac.uk/\lb{2}$\sim$guy/.  
Radio data at 15 GHz were observed    
with the Ryle Telescope, Cambridge (see Pooley \& Fender, 1997, for   
a description of the techniques used); X-ray data (ASM flux and hardness    
ratio) are taken from the {\it Rossi}-XTE quick-look web pages:  
http://\lb{2}xte.mit.edu/\lb{2}ASM\_lc.html. The approximate    
dates of OBS1 and OBS2 are shaded. Note the anti-correlation    
between X and radio. }   
\hfill   
\label{fig:radio}   
\end{figure}   
   
        \section{Observations and data reduction}   
        \label{sec:reduction}   
   
To trigger the ToO observation, we set a condition on    
variability ($rms$ single-dwell points of {\it Rossi}-XTE ASM    
to stay $< 9$\% within one day) and a condition on flux    
({\it Rossi}-XTE ASM daily rate below a threshold of $30$    
cts/s).\footnote{ASM is the {\it All Sky Monitor} onboard {\it Rossi}-XTE.   
We used the public ASM ``Weather Map'' mask, accessible at:   
{\sf http://\lb{2}heasarc.gsfc.nasa.gov/\lb{2}xte\_weather/}.}    
Both triggering conditions were met during the long ``plateau",    
a lower-flux state also characterized by a more intense radio emission   
(Foster \ea\ 1996),   
which lasted several weeks during the Spring of 2004 (Fig.~\ref{fig:radio}).   
The observation could be therefore started, in EPIC timing mode,    
on April 17, 2004 (hereafter: {\it OBS1}).   
   
However, the source counts were enough to saturate the telemetry,   
even after switching off the MOS and RGS. Thus a second   
part of the observation was performed on April 21, 2004 (hereafter:   
{\it OBS2}), this time in (EPIC-pn) burst mode.    
   
Given the typical plateau behavior, with strong jet activity and high radio level,    
we conclude that the source was caught in the conventional ``hard'' (C) spectral   
state and $\chi$ variability class, as defined by Belloni \ea\ (1997a,b, 2000),    
in both observations. This is similar to the condition in which the source was seen    
by simultaneous {\it INTEGRAL} (rev.57) and {\it Rossi}-XTE observations  
performed in the   
Spring of 2003 (Hannikainen \ea\ 2003, 2004, 2005; see Fuchs \ea\ 2003 for a    
multi-wavelength study of one of these observations).\footnote{This radio-loud    
state had been described already    
by Muno \ea\ 2001, and by Trudolyubov 2001 (as ``type II'').}    
    
Count rate thresholds for pileup in timing modes are reported in    
the XMM User's Handbook.\footnote{See Table 3 in: Ehle \ea\ (2004) -- {\sf    
http://\lb{2}xmm.vilspa.esa.es/\lb{2}external/\lb{2}xmm\_user\_support/\lb{2}documentation/\lb{2}uhb/\lb{2}node28.html\#2828 }}    
The MOS count rate registered in both observations   
($\sim 130$ and $160$ cts/s) 
is well above the quoted limit (100); therefore EPIC-MOS   
data are unusable,\footnote{Excising    
the core of the PSF is enough to cope with this problem in the case of imaging    
modes, but it is not applicable for timing modes.} and we restricted the analysis    
to EPIC-pn and RGS data (the latter are available only for OBS2).   
   
Since in this state the source is stable, we can use the whole integration   
time to obtain very high statistics and signal/noise.   
Through adopting the customary method of extracting high-energy    
($E > 10$ keV) background light curves, we checked for intervals of low    
contamination in both observations. While no background flaring is apparent during OBS1,    
we did initially select a 9~ks low-background interval ($198951000\div198960000$ s)    
in OBS2. We performed preliminary OBS2 data analysis with this selected   
dataset, then afterwards verified that the results were not modified when    
considering the whole observing interval. Therefore, in the following, we report   
results of the data analysis that was performed using {\it all}    
observing time, e.g. with the highest possible statistics.   
This also allowed us to refine the spectro-variability    
analysis (e.g. the analysis of root mean squared variability vs. energy:    
Ponti \ea\ 2004) as much as possible, which we    
performed to possibly disentangle the various contributions   
to the observed spectra and, thus, to get an independent validation   
of the adopted emission/absorption model.   
The high number of counts allows a high-resolution analysis of   
$rms$ vs. $E$ over the   
whole EPIC-pn energy band (about 35 energy bins for OBS2, and several   
dozen energy bins for OBS1; the minimum timescales of variation we    
could consider were, respectively, $\sim 1000$ and $\sim 100$ s).   
We used background-subtracted light curves in order to avoid contamination    
at the lowest and highest energies ($E \lesssim\ 2$ and $E \gtrsim\ 7$   
keV),  
and found a mean $rms$ value on the order of 5\% ($0.02\div0.07$) in both    
observations, compatible with a null $rms$ when Poisson noise is removed,   
with no visible features in the ``$rms$ spectra''. We further verified this    
by studying $F_{\rm var}$ vs.    
$E$.\footnote{The nominal time resolution is 29 and 7 microseconds for EPIC-pn    
timing and burst modes, respectively. However, all results derived from    
the light curves must be taken with caution:    
in the first observation, even after removing   
all telemetry drops, the collected photons ``saturated''   
the instrument's bandwidth; therefore, energy-dependent oscillations    
in the count rate may have been ``cut off''. 
On the other hand, photons collected in the second observation, i.e.   
in burst mode, are selected on the basis of an {\it ad hoc} duty cycle in the    
reading process (events are not registered for 97\% of the time). }   
   
The data were reduced with the standard SAS v6.1 and FTOOLS   
software packages. For EPIC-pn,   
single and double events were used in the timing modes ({\sc PATTERN} $<= 4$);   
the fit was restricted to energies $> 0.5$ ($> 0.4$) keV in timing (burst)    
mode in order to avoid the increased noise.   
As recommended by the XMM-{\it Newton} Helpdesk, no selection on the RAWY coordinate    
was applied since this is related to a fine-time and the selection would    
therefore exclude certain time periods. However,  RAWY $< 160$ was imposed   
as a condition in burst mode to avoid direct illumination by the source.   
Response and auxiliary matrices were created with the SAS tools   
{\sf rmfgen, arfgen, rgsrmfgen}.   
   
For the fit procedure we    
assumed a source distance of 12.5 kpc and inclination $i \sim 70^{\rm o}$    
(Rodriguez \& Mirabel, 1999; Fender \ea, 1999).

\subsection{First observation }   
   
The April 17 observation (OBS1: XMM-{\it Newton} rev.798,    
start time: 14:18:56 [JD2453113.096], stop time: 20:03:37 [JD2453113.336]) was    
performed in (both pn and MOS1) timing mode, as requested in the    
AO2 proposal mainly with the aim of avoiding pile-up problems. Moreover,    
the MOS2 camera and both RGS had to be switched off, to avoid overloading the   
telemetry.    
Nevertheless, the source flux was still so high that telemetry drops were    
frequent even in pn data, resulting in short gaps in the light curves,    
lasting a few to several seconds, for a total of about 75\% of the    
observing time. They cause the search for (quasi-) periodic    
oscillations to be senseless on $\sim 1$ Hz timescales.    
Source light curves do not show any appreciable variations at timescales    
$\sim$ {\it a few} to $\sim 100$ s; however, as explained above (see Footnote 6),   
due to the full scientific buffer and telemetry drops, this result should    
be taken with caution.   
   
Even though   
affected by telemetry drops for up to about 75\% of the observing time, EPIC-pn   
data are well below the pile-up count rate threshold\footnote{For EPIC-pn    
timing mode the pile-up threshold for a point source is 1500 cts/s,    
the telemetry threshold is 300 cts/s, while we registered $\sim 650$ cts/s. }    
and can therefore be safely used for spectral analysis.   
   
The data were selected from a four-column strip centered on the source   
throughout the timing-mode (one-dimensional) ``image''; a similar   
strip was selected far away from the source as a background sample.

\subsection{Second observation }   
   
While the source essentially persisted in the same state, a second    
observation was started on April 21st (OBS2, rev. 800,    
start time: 13:49:23 [JD2453117.076], stop time: 20:56:55 [JD2453117.373]). To cope    
with the telemetry problem, this time we used EPIC-pn in burst mode, MOS in    
timing mode, and RGS in high count rate (HCR) mode.   
The other cameras were off.    
   
In this observation, a background flare is apparent at the very beginning   
both in EPIC and RGS data; however, due to the very high S/N and shortness   
of the flare, it does not really affect the spectra.   
Source data were selected from a six-column strip (i.e. two columns   
more than in timing mode, in order to increase the statistics)
throughout the burst-mode (one-dimensional) ``image''; a similar   
strip was selected, far away from the source, as a background sample.   
   

\begin{figure}   
\hspace{-1.0cm}   
\includegraphics[angle=270,width=0.55\textwidth]{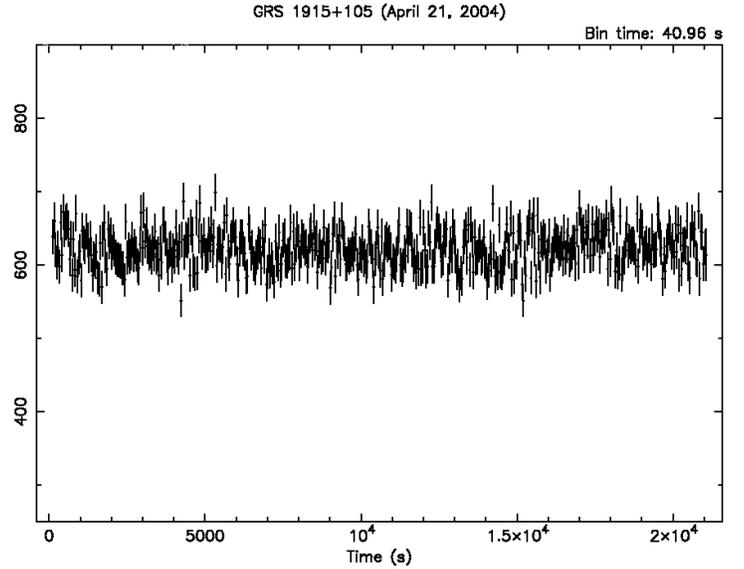}      
\caption{The source light curve in the 0.4 -- 10.0 keV band    
during OBS2 (EPIC-pn, burst mode, {\it not} background-subtracted).    
The whole 20946 s observation has a mean 618 (i.e. $18.54/0.03$) cts/s;    
the background has just about 18 (i.e. $0.5320/0.03$) cts/s. }   
\label{figps}   
\end{figure}   
   
\begin{figure}   
\hspace{-0.5cm}   
\includegraphics[angle=0,width=0.55\textwidth]{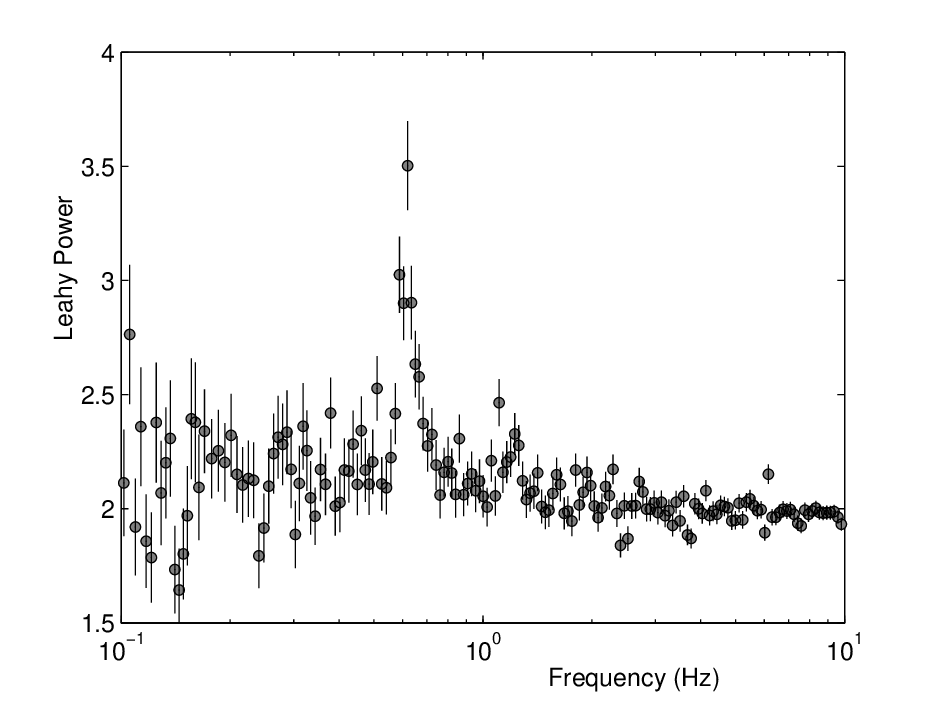}      
\caption{Power spectrum in the 0.4 -- 13.0 keV band of the OBS2 (April 21)     
EPIC-pn data. }   
\label{figqpo}   
\end{figure}

The EPIC-pn data are affected neither by telemetry drops nor by pile-up, in this    
observation, and can therefore be used for both spectral and timing analyses.   
However, as far as timing/variability studies are concerned, some caution is   
necessary since burst-mode data suffer from the intrinsic photon loss related   
to the 3\% duty cycle (cp. Footnote 6).   
While the light curves only show tiny variations (up to a few percent,    
but compatible with zero: see Fig.   
\ref{figps}) at all timescales bigger than a few seconds, a quasi periodic  
oscillation (QPO) at about   
0.6 Hz, with a possibly harmonic signal at 1.2 Hz, is seen in the EPIC-pn data    
collected in burst mode (Fig.~\ref{figqpo}). This frequency is    
expected in $\chi$ variability modes of   
GRS\,1915+105, when the phenomenological correlation   
with the source flux is taken into account (c.p. Fender \& Belloni 2004).   
A forrest of narrow peaks at $\nu > 100$ Hz is apparent, due to    
spurious signals of instrumental origin.    
   
\subsubsection{The RGS data}   
  
\begin{figure}   
\centering   
\hfill   
\includegraphics[angle=-90,width=0.48\textwidth]{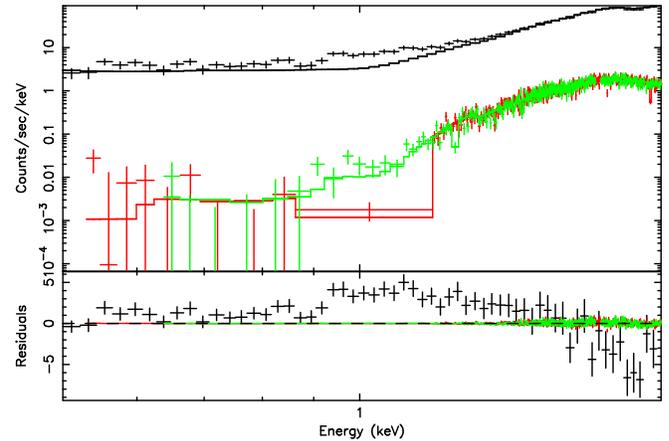}   
\caption{The RGS and pn (burst mode) spectra fitted in the 0.5-2 keV range  
with a simple absorbed power law. The RGS and pn spectra clearly disagree each other.}   
\hfill   
\label{fig:rgs}   
\end{figure}

Thanks to their excellent energy resolution,   
RGS cameras should be able to reveal subtle details  in the 0.3--2.0 keV    
band. However, the source is so strongly absorbed below $1\div2$ keV that    
these data do not actually add much information. 
We used them mainly to check the EPIC-pn spectrum: 
in Fig.~\ref{fig:rgs} we show the 0.5-2 keV spectra of both instruments,  
fitted with a simple absorbed power law model, with  
$\Gamma$ fixed to 2 (see Sect. 3.2).\footnote{Multiplicative 
constants to the RGS have also been added to allow for 
intercalibration problems (the best fit values for the constants are 1.28 
and 1.22 for the RGS1 and RGS2, respectively). RGS data,
collected in the standard CCD read-out scheme, are marginally
affected by pile-up, which amounts to a few percent and mainly affects 
the energies where the spectrum peaks. }
The RGS and EPIC-pn spectra are clearly different, especially with regard
to the broad excess around 1 keV in the latter spectrum. The RGS spectrum
shows a fast decline (more than two orders of magnitude, with very few 
counts at the lower energies), without any apparent features. 
In Sect. 3.1.5 we go through the alternative hypotheses 
that can be invoked to explain the RGS--pn discrepancy.

RGS normally have a 15 ms timing resolution in HTR mode; we produced standard    
light curves, but did not get any useful results due to the low statistics
on most of the band. 
  
        \section{Spectral analysis}   
        \label{sec:spectral}   
   
At the flux level of GRS~1915+105 ($\sim 10^{-8}$ erg cm$^{-2}$ s$^{-1}$),   
it is very likely that systematic errors cannot always be neglected    
with respect to the statistical ones. However,    
introducing an energy-independent systematic uncertainty would    
prevent a meaningful use of the $\chi^2$ statistics.   
Features at the 1-2\% level must therefore be taken with caution as   
possibly of instrumental origin (cp. Kirsch 2005).   
   
Even though the background issue is not relevant on most of the   
energy band, due to the extremely high S/N, nevertheless it may play   
a role at the highest energies 
($E \gtrsim\ 10$ keV), as well as at the   
energies where the source is strongly absorbed 
($E \lesssim\ 1.5$ keV).  
We therefore performed background subtraction; all reported results   
refer to background-subtracted analysis, unless stated otherwise.   
Of course, if a dust halo is present 
(cannot be checked with our data, given the lack of imaging capabilities 
of the timing modes; but see Greiner et al., 1998, on the issue), 
background subtraction may cause a spectral distortion in the soft   
band, depending on the distance from the source at which the background   
spectrum is taken (see below, Sect. \ref{sec:exc}). 
   
Spectra were rebinned in order to have about three bins per energy resolution   
element and at least 20 counts per bin at the same time.    
Spectral fits were performed with the {\sc XSPEC} v.11.3.0 software    
package. In the following, all errors refer to a 90\% confidence level    
for one interesting parameter ($\Delta\chi^2$=2.71).

        \subsection{The timing-mode observation}   
   
As discussed in the previous section,   
we are confident that the telemetry problems do not affect the spectrum
significantly.
Given the much larger total number of counts in OBS1, we decided to    
use it first for the spectral analysis.    
   
\begin{figure}   
\epsfig{figure=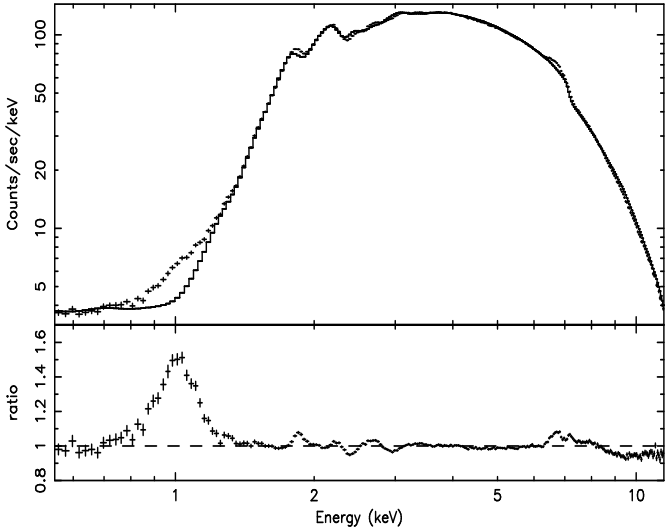,height=8.8cm,width=7.0cm,angle=-90}   
\caption{The OBS1 spectrum and data/model ratio when fitted with a simple power law and   
cold absorption with variable abundances.}   
\label{gio1pl}   
\end{figure}

To illustrate the spectral complexity,   
in Fig.~\ref{gio1pl} we show the best fit spectrum and data/model ratio after   
fitting with a simple power law absorbed by cold matter. Element abundances were   
left free to vary with respect to the solar value ratios. Several features are    
apparent: an excess around 1 keV, several wiggles between $\sim 1.5$ and 3 keV,   
an excess at the energies of the iron line complex, and a deficit above 8 keV.   
   
\begin{figure}   
\epsfig{figure=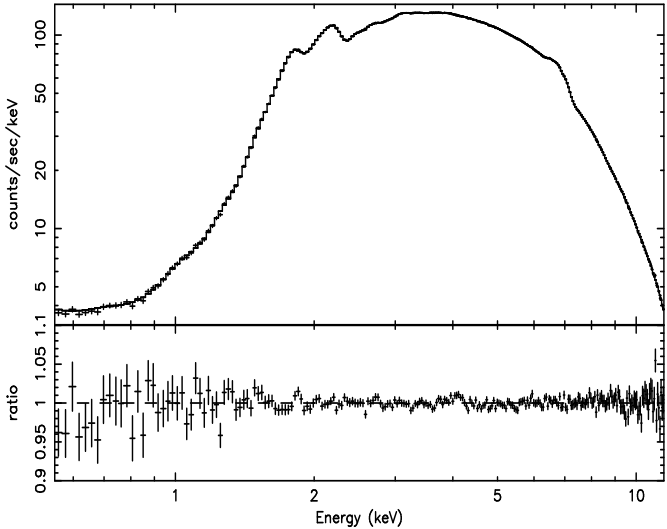,height=8.8cm,width=7.0cm,angle=-90}   
\caption{The OBS1 spectrum and data/model ratio for the best fit model (see text and    
Table~\ref{bestfittable}).}   
\label{bestfit}   
\end{figure}   
   
After several attempts, we found a -- both statistically ($\chi^2$/d.o.f.=317.5/227)   
and physically -- reasonable description  of the spectrum   
in terms of (see Table~\ref{bestfittable}, and Fig.~\ref{bestfit})    
a power law with $\Gamma\sim$1.7,   
several absorption and emission features,    
an ionized and optically thick reflection component (and/or a warm absorber),    
and an optically thin reflection component  to model the 1 keV excess.    
Given the fact   
that all remaining deviations of the data from the model are no larger than    
about 1-2\%, i.e.   
on the order of the systematic errors in the calibration (which, for most of the energy   
range under analysis, is larger than the statistical errors), we did not try to  
improve the fit further.     
   
We now discuss  the various spectral components separately for the sake of clarity.   
   
\subsubsection{The continuum}   
\label{sec:cont}   
   
First of all, we note that a simple power law is basically sufficient to describe    
the continuum, apart from an additional component required to fit the $>$8 keV    
deficit and an apparent broad excess around $\sim 1$ keV.   
Substituting the power law with simple Comptonization models (like {\sc comptt}    
or {\sc compst}) does not lead to any significant improvement in the fit.   
No thermal emission from an accretion disk is   
apparent (the inclusion of a multicolor disk component is not required by    
the data). Of course,    
this strongly suggests that the innermost disk regions were absent during the    
observation. The apparent 1 keV excess is 
far too narrow to be fitted by even a single temperature blackbody  
(i.e. the narrower among physically motivated continuum models)  
or to be residual from ``distorted'' disk thermal emission.    
The nature of this feature will be discussed separately in Sect.~\ref{sec:exc}.   
   
\subsubsection{The cold absorber}   
\label{absorb}   
   
The photoelectric absorption has been fitted with the {\sc XSPEC} model {\sc varabs}, which    
assumes neutral matter and allows the elemental abundancies to vary separately.    
In practice, for the sake of simplicity, we grouped the elements so that within each   
group the elements are linked as: H, He, C, N, O;    
Ne, Na; Mg, Al; Si; S; Cl, Ar, Ca, Cr; Fe, Co, Ni.  The rationale behind this    
choice is both physical and practical. In fact, we grouped elements   
together whose origin is very likely common, but also those which are less    
abundant (and therefore cannot be measured easily independently) 
with very abundant ones: e.g. Co and Ni with Fe.   
We also grouped H and He with C, N, and O because their effect    
is small in our energy range.   
   
In Table~\ref{bestfittable} the best fit values are reported. They are given in units of    
hydrogen equivalent column densities for solar abundances as in Anders \&    
Grevesse (1989).   
A significant overabundance of the heavier elements with respect to the lighter ones   
is apparent. These results are only partly consistent with those derived from the    
{\it Chandra} HETG observation (Lee et al. 2002).    
In particular, Fe and S abundances are roughly the same, while   
our values for Mg are about twice, and 2/3 for Si. Fixing the abundances to the values of   
Lee et al. (2002), the fit is unacceptable  ($\chi^2$/d.o.f.=659.9/231).   
We note that our observation has a lower energy resolution, but many more photons.    
   
These overabundancies with respect to the solar values (the differences are even greater   
assuming the ISM values, see Wilms et al. 2000) suggest that   
a significant fraction of the absorber is local to the source.    
The value for the low Z elements (i.e. 1.6$\times10^{22}$ cm$^{-2}$)   
can then be assumed as the upper limit to the ISM absorber   
column density. It is interesting to note that this value is similar to   
the one estimated by fitting the $\sim$1 keV excess (see below).   
   
\subsubsection{Reflection or warm absorption?}   
\label{reflec}   
   
The deficit above 8 keV can be fitted equally well by a simple smeared    
edge model  ({\sc smedge}) or by an ionized and relativistically blurred    
reflection component ({\sc refsch}). As the smearing edge is not physically    
meaningful {\sl per se}, but is usually assumed to represent the reflection    
component, we will not discuss the fits with this component further.    
   
The model adopted for the disk reflection component includes the relativistic    
effects in Schwarzschild  metric. The inner and outer radii are linked to those    
of the H--like iron line (see below). The disk inclination angle has been    
fixed to 70$^{\circ}$, and the matter temperature to $10^6$ K.   
The very large value of the inner radius (about 320$r_g$) justifies   
{\sl a posteriori} the use of the Schwarzschild metric instead of the Kerr one    
(differences between the two metrics, even for a maximally rotating BH, are    
insignificant above $\sim$10$r_g$), and it agrees with the lack of any detectable    
thermal disk emission.   
The iron abundance is five times the solar value, roughly consistent with    
the overabundance   
found in the cold absorber (see above). The ionization parameter is pretty large,    
in agreement with the presence of He-- and H--like iron lines. The significant    
solid angle subtended by the disk to the illuminating source, i.e.    
0.3$\times$2$\pi$, implies that the emission occurs at distances comparable to   
the inner disk radius.   
The adopted model is necessarily a simplification of the real situation,    
as it adapts a unique ionization for the whole disk, while the ionization very likely    
decreases with increasing radius. The derived parameters, therefore, must be    
taken with caution and considered only as indicative.   
   
The presence of a H--like iron absorption line (see below) suggests    
warm absorption.   
Substituting the reflection component with a warm absorber (model {\sc absori}),    
a significantly worse, but not completely unacceptable fit is found    
($\chi^2$/d.o.f.=384.5/226), with a comparable iron overabundance    
($A_{\rm Fe}$=6.6) and power law photon index ($\Gamma\sim1.69$).   
However, the column density ($\sim 7.7\times10^{21}$   
cm$^{-2}$) is larger than that implied by   
the value of the EW of the absorption line (see Table~\ref{bestfittable}),    
given the large iron overabundance (e.g. Bianchi \ea\ 2005). Moreover,   
the presence of ionized disk lines (see below) {\it requires}    
the presence of the ionized reflection component. We thus fitted the spectrum    
with both the reflection and the warm absorber. No significant improvement is    
found with respect to the fit with the reflection only, but the upper   
limit to the column density, $\sim$2$\times10^{21}$ cm$^{-2}$,    
is now consistent with the absorption line EW.   
  
   
\begin{table}[t]   
\begin{center}   
\begin{tabular}{|c|c|}   
\hline   
& ~ \cr   
& {\bf Power law} \cr   
$\Gamma$ & 1.686$^{+0.008}_{-0.012}$ \cr   
& ~ \cr   
& {\bf Cold absorption} \cr   
$N_{\rm H,He,C,N,O}$ & 1.60$^{+0.17}_{-0.29}\times10^{22}$ cm$^{-2}$\cr   
$N_{\rm Ne,Na}$ & 7.46$^{+0.31}_{-0.72}\times10^{22}$ cm$^{-2}$ \cr   
$N_{\rm Mg,Al}$ & 7.57$^{+0.54}_{-0.16}\times10^{22}$ cm$^{-2}$ \cr   
$N_{\rm Si}$ & 5.70$^{+0.07}_{-0.12}\times10^{22}$ cm$^{-2}$ \cr   
$N_{\rm S}$ & 4.69$^{+0.07}_{-0.69}\times10^{22}$ cm$^{-2}$ \cr   
$N_{\rm Cl,Ar,Ca,Cr}$ & 11.0$^{+1.3}_{-1.6}\times10^{22}$ cm$^{-2}$ \cr   
$N_{\rm Fe,Co,Ni}$ & 11.7$^{+0.2}_{-0.2}\times10^{22}$ cm$^{-2}$ \cr   
& ~ \cr   
& {\bf Emis. lines} \cr   
$E_l$ & 1.846$^{+0.006}_{-0.005}$ keV \cr   
$F_l$ & 1.35$^{+0.15}_{-0.07}\times10^{-2}$ ph cm$^{-2}$ s$^{-1}$ \cr   
$EW$  &  19 eV \cr    
& ~ \cr   
$E_l$ & 2.244$^{+0.007}_{-0.010}$ keV \cr   
$F_l$ & 3.62$^{+0.46}_{-0.62}\times10^{-3}$ ph cm$^{-2}$ s$^{-1}$ \cr   
$EW$  &  8 eV \cr   
& ~ \cr    
& {\bf Abs. line} \cr   
$E_l$ & 6.95$^{+0.01}_{-0.03}$ keV \cr   
$F_l$ & -7.9$^{+0.2}_{-0.1}\times10^{-4}$ ph cm$^{-2}$ s$^{-1}$ \cr   
$EW$  & -9 eV \cr    
$\sigma$ & 1 eV (frozen) \cr   
& ~ \cr   
& {\bf 1st disk line (He-like Fe)} \cr   
$E_c$ & 6.7 keV (fixed) \cr   
$r_i/r_g$ & 580$^{+210}_{-120}$ \cr   
$F_l$ & 2.37$^{+0.18}_{-0.18}\times10^{-3}$ ph cm$^{-2}$ s$^{-1}$ \cr   
$EW$ & 28 eV \cr   
& ~ \cr   
& {\bf 2nd disk line (H-like Fe)} \cr   
$E_c$ & 6.96 keV (fixed) \cr   
$r_i/r_g$ & 320$^{+80}_{-60}$ \cr   
$F_l$ & 2.20$^{+0.19}_{-0.21}\times10^{-3}$ ph cm$^{-2}$ s$^{-1}$ \cr   
$EW$ & 28 eV \cr   
& ~ \cr   
& {\bf Reflection} \cr   
$R/2\pi$ & 0.35$^{+0.02}_{-0.02}$ \cr   
$A_{\rm Fe}$ & 5.2$^{+0.7}_{-1.9}$ \cr   
$\xi$ & 940$^{+190}_{-80}$ erg cm s$^{-1}$ \cr   
$r_i/r_g$ & 320$^{+80}_{-60}$ \cr   
& ~ \cr   
$F_{\rm 2-10~keV}$ &   
$\sim 0.6$ (unabs: $\sim 0.87$) $10^{-8}$ erg cm$^{-2}$ s$^{-1}$ \cr  
& ~ \cr  
$\chi^2$/d.o.f. & 317.5/227\cr   
& ~ \cr   
\hline   
\end{tabular}   
\caption{The best fit model parameters for the timing-mode observations.   
The column densities are the Hydrogen equivalent column for solar abundances    
as in Anders \& Grevesse (1989).   
The outer radii of the reflection and disk line models have been linked together in the   
fit, yielding $r_o/r_g = 900^{+190}_{-90}$; $r_i$ in the 2nd disk line model has been    
linked together to $r_i$ of the reflection component.   
See the text for further discussion. }   
\label{bestfittable}   
\end{center}   
\end{table}

\subsubsection{The iron lines}   
\label{sec:felines}   
   
In Fig.~\ref{gio1pl}, a clear broad excess at the energy of the iron line    
complex is apparent. The excess is at larger energies than 6.4 keV, strongly    
suggesting ionized iron. We fitted this excess with two unresolved emission    
lines ($\sigma$=10 eV) peaking at 6.7 keV and 6.96 keV, which correspond to He--    
and H--like iron, respectively; and after Lee \ea\ (2002) we also included   
an unresolved absorption line with energy free to vary around the value   
of the H--like iron.  The fit is unacceptable ($\chi^2$/d.o.f.=617.2/225), and not   
surprisingly, the absorption line was not found. A much better, but still hardly   
acceptable, fit ($\chi^2$/d.o.f.=425.6/223) is found when leaving the widths of    
the lines free to vary. We found $\sigma$=129 eV and 174 eV for the He-- and    
H--like lines, respectively;    
an absorption line centered at $\sim 6.95$ keV is also found.   
The high values of $\sigma$, along with   
the likely presence of the disk reflection component, strongly suggest   
that the lines are relativistically broadened: we therefore substituted the    
broad Gaussians with {\sc diskline} models, with the   
outer radius linked to that of the reflection component, as well    
as the inner radius of the H--like line (the inner radius of the He--like left    
free to vary, in the assumption that the matter is more ionized in the innermost    
regions). The results are reported in Table~\ref{bestfittable}.    
The fit is significantly better than with the Gaussians and -- as expected on    
both physical grounds and from the fact that the width of the He--like line is    
smaller -- the inner radius for this line is larger than that of the H--like    
line (the values of the radius must of course be taken merely as indicative,    
see discussion above). The energy of the narrow absorption line,   
6.95$^{+0.01}_{-0.03}$ keV, is just below the rest frame   
value.\footnote{Actually,    
this is a blend of two lines, with centroid energy at 6.966 keV.}   
The line was assumed to be narrow ($\sigma \equiv 1$ eV); when leaving    
$\sigma$ free to vary, no improvement in the fit is found.    
   
We  also tried to fit the iron emission with just one disk line model, with the energy    
free to vary. The fit is of comparable statistical quality, but the line    
energy, 6.86$\pm$0.01 keV, does not correspond to any rest frame atomic energy   
of significance. Moreover, the good fit is obtained at the expense of a very    
large value for the iron abundance of the reflection component ($\sim$18 times    
the solar value). Therefore, the model with two disk lines has to be preferred.   
   
The EW of both lines (28 eV) seems rather small when compared to the amount of    
the reflection component (Matt \ea\ 1996). However, one must note again that the    
ionization most likely changes along the radius in the disk. While reflection occurs    
(even if with changing spectra) throughout the disk, line emission    
occurs only if the corresponding ion is present, which probably only happens    
in a fraction of the disk.

\subsubsection{The $\sim$1 keV excess}   
\label{sec:exc}   
  
In Sect. 2.2.1 we have shown that, during the burst-mode observation, the RGS 
and pn disagree, with the former not showing the 1 keV excess.   
(Unfortunately, during the timing-mode observation the RGS were switched  
off). There are several possible explanations for both the excess and the   
RGS-pn discrepancy.
As we will see, none of them is fully satisfactory, and we consider the issue still 
open as we wait for improvements in calibrations.  
   
A pn instrumental origin for the broad excess seems unlikely    
for various reasons. The amplitude    
(about 40\% of the underlying continuum, see the data vs. model ratio    
in Fig.~\ref{gio1pl}) is far too large, given the known   
calibration uncertainties (cp. Kirsch 2005).    
Moreover, the excess is present in the Timing as well as in the burst mode, and    
was also observed with {\it Beppo}SAX (cp. Fig.~3 in: Feroci \ea\ 1998).   
A similar feature has also been seen in other binaries (see e.g. the case of the    
dipping LMXB 4U\,1323-62: Boirin \ea\ 2005), but not in all; in particular, it is  
not present in the burst mode observation of the Crab Nebula (Kirsch et al. 2005). 

However, given the unusual brightness of our source and the fact that the excess 
arises in the part of the spectrum where the count rate drops precipitously due 
to absorption, a calibration problem that is possibly related to the redistribution matrix 
cannot be completely excluded at the moment. \\ 
  
Let us then assume that this excess is real. We could not fit this feature with    
any continuum model put behind the whole absorber. As seen in    
Sect.~\ref{sec:cont}, it cannot be fitted with blackbody models; the only model 
able to fit it is a broad Gaussian with an EW of several tens of keV, which has no    
physical explanation. Good fits are instead found if the emission is screened    
by a thinner absorber.  
There are three possibilities in this respect: (i) a    
confusing source, (ii) a dust scattering halo, or (iii) a component    
related to the system but lying outside the intrinsic absorber.  
   
The confusing source hypothesis seems unlikely.   
In timing modes, source data are extracted from a narrow strip (here about 16'');   
the required, ``polluting'' source would have to be bright (0.5-2 keV  
observed flux of about  4$\times$10$^{-12}$ erg cm$^{-2}$ s$^{-1}$): but   
no source is present in the {\it ROSAT} All Sky Survey   
within a radius of 0.3$^{\circ}$ around GRS\,1915+105 (we checked with the HEASARC  
tool {\sc PIMMS} that such a source would be well above the typical RASS detection limit;  
of course, the possibility of a variable source cannot be ruled out).
   
Since the dust scattering is energy dependent,   
ignoring a possible halo component leads to a distortion of the spectrum   
in a small energy band, because of an incorrect background subtraction   
(depending on the distance from the source at which the background   
spectrum is taken); however, less soft photons would be "observed".  
A possibility is that we are including the halo in the extraction region.  
Since there is no obvious dust scattering model to fit the data,   
imaging may be the only way to test the halo hypothesis: given the particular modes  
adopted, this sort of analysis is very difficult (and altogether impossible 
in the burst mode). We only note that the dust scattering  
halo observed in Cyg\,X-2 (Costantini et al. 2005) has quite a broad spectrum 
(when normalized to the source spectrum), broader than the feature under discussion.  
Another problem with the scattering halo hypothesis is that the  
relative contribution of the excess is similar in the timing- and burst-mode   
observations, which is possible only if most of the halo is within 16'', i.e.  
the width of the strip from which the timing-mode spectrum was extracted.   
  
Even if unsatisfactory in many respects, the two above-mentioned hypotheses have   
the merit of invoking extended or off-axis emission, which can in principle help  
explain the presence of the excess in the pn spectra but not in the RGS. However,  
as the RGS spectrum is extracted over a region of about 2.5 arcminutes, the halo   
hypothesis seems to be excluded for the reasons discussed above. A confusing 
source, provided it lies at a greater distance, is instead still tenable.  
We note that all {\it Chandra} observations of GRS\,1915+105 have been performed  
in the CC (i.e. non imaging) mode, therefore the above-mentioned hypotheses 
could not be tested yet. \\
  
In the assumption that the RGS spectrum 
suffers from instrumental/calibration problems,   
the 1 keV component could be related to the system, and a possible   
candidate would be the emission by a disk wind. 
Both a collisionally ionized plasma model   
and a photoionized plasma model do indeed fit the EPIC-pn data well.    
   
In the first case, the temperature of the plasma is about 1.3 keV (obtained using    
the {\sc mekal} model), with metal abundance 1.2 times the solar value; the ISM    
absorber has an $N_{\rm H}$ of 9$\times10^{21}$ cm$^{-2}$. The emission integral   
is 7.2$\times10^{57} d^2_{10}$ cm$^{-3}$ (where $d_{10}$ is the distance    
to the source in units of 10 kpc), which, assuming that the density of the matter    
is constant along the emitting region, translates into a mass of the   
wind $M_w \approx$ 6 $M_{\odot} d^2_{10} n^{-1}$, where $n$ is the number density.    
Assuming that the size of the disk wind is   
at most on the order of the binary separation   
(about 5$\times10^{10}$ cm according the binary system parameters given   
by Harlaftis \& Greiner 2004), we obtain a lower limit to $n$ of about    
8$\times10^{12} d_{10}$ cm$^{-3}$. With this size, the Thomson optical depth    
would already be substantial ($\sim 0.25$), and be increasing further with decreasing    
size. Thus, the matter would not be really optically thin, so this solution is    
not self-consistent.   
   
Alternatively, the matter may be in photoionization equilibrium with the   
emission due to reflection of the primary X--ray radiation. To model this    
case, we adopted a power law with the same photon index as the primary one (to    
model Thomson scattered radiation) plus a Gaussian line. In this case, the ISM    
$N_{\rm H}$ is larger, $\sim$1.5$\times10^{22}$ cm$^{-2}$ (this value may be assumed    
as an estimate of the ISM column density)\footnote{For the sake of simplicity we fitted   
the absorption with a pure gas photoelectric absorption (model {\sc wabs} in {\sc XSPEC}).  
Using a more appropriate ISM model that includes the effects of dust 
({\sc TBabs} in {\sc XSPEC}, Wilms et al. 2000) does not change the results significantly. 
The fit returns basically the same $\chi^2$ and the same column density.}. 
From the observed flux, the optical depth of the    
wind is about 0.001/$f$, where $f$ is the covering factor of the wind to the    
primary radiation. The centroid energy of the Gaussian line ($\sim$0.97 keV), its    
width (90 eV), and its EW against the reflected continuum (5.6 keV), all suggest a    
blend of Ne K and Fe L lines.

\subsubsection{Other emission/absorption lines}   
   
Other emission and absoption lines are required to fit the spectrum. They   
are: two unresolved emission lines at about 1.85 keV and 2.25 keV, with    
EW=19 eV and 8 eV, respectively; two broad absorption lines at about 2.5    
and 2.93 keV, with $\sigma$ of 0.11 and 0.14 keV and EWs of $-17$ and    
$-30$ eV, respectively.    
   
Some of these features may be instrumental artifacts, since this energy range is    
notoriously troublesome due to the changes in the effective area.\footnote{See   
Figs.30 and 31 in the XMM-{\it Newton} User's Handbook:    
{\sf http://\lb{2}xmm.vilspa.esa.es/\lb{2}external/\lb{2}xmm\_user\_support/\lb{2}documentation/\lb{2}uhb/\lb{2}node32.html} .} However, some of them appear too   
prominent to be entirely due to calibration problems. The first emission line can be    
readily identified with the Si {\sc xiii} recombination line. If arising   
from the same photoionized material responsible for the $\sim$1 keV excess (see above),   
its total EW would be about 16 keV, a very large but not fully unrealistic value, given    
the large overabundance we found in the local absorber and provided that the   
reflecting matter is still optically thin to resonant scattering (Matt et al. 1996),    
which is possible if the matter is largely turbulent. The same is true for the second   
emission line, which can be identified with Si {\sc xiv} and S {\sc xii}, even if the    
best fit value of the energy is slightly lower.   
   
More difficult  to explain are the two broad absorption lines. The energies correspond   
to the K$\alpha$ and K$\beta$  recombination lines from S {\sc xv}.   
In this case, the absorbing material cannot be the same as that responsible for the    
Fe {\sc xxvi} absorption line (see Sect.~\ref{sec:felines}), since the latter    
is more ionized.    
It is also worth recalling that the modeling of the absorption edges is oversimplified,   
because it does not take into account X-ray absorption fine structures (XAFS),    
which are generated if the material is   
in grains (e.g. Lee et al. 2002); this may introduce spurious features in the    
residuals, even if it cannot fully explain them. Moreover, we again stress   
that it is  not inconceivable that a significant part of these features    
are of instrumental origin. Data with higher energy resolution and    
optimized calibration are necessary to solve this issue.

\subsection{The burst-mode observation}   
\label{sec:burst}   
   
We fitted the burst-mode spectrum with the same model used for the timing-mode    
spectrum, fixing those parameters that are not expected to vary.    
In particular, we fixed the interstellar absorption column density, the ratio    
between the elemental    
abundancies of the local absorber, and the iron abundance of the reflection    
component at the best fit values obtained for the timing-mode observation.   
A good fit ($\chi^2$/d.o.f.=248.2/219) is found. The best fit parameters   
are summarized in Table~\ref{bestfittable_2}.   
   
\begin{figure}   
\epsfig{figure=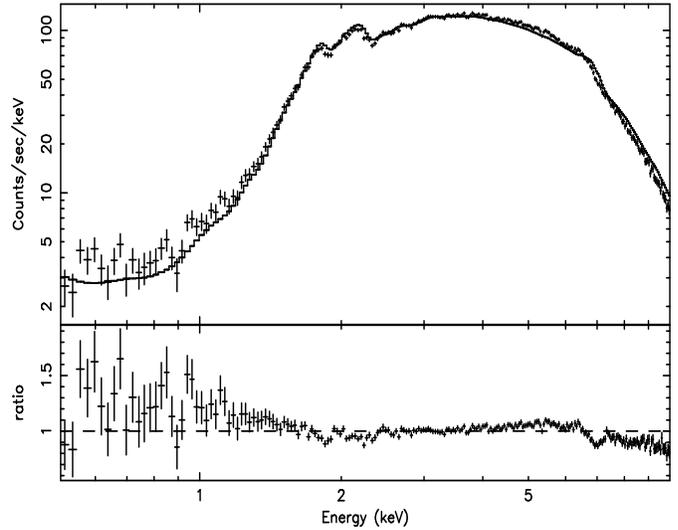,height=8.8cm,width=7.0cm,angle=-90}   
\caption{The spectrum and data/model ratio of the burst-mode observation    
with the best fit model for the timing-mode observation. Clearly, iron line   
emission is over-subtracted, and  
the deficit at higher energies turns out to be more important than in OBS1;  
note again the strong excess at $\sim$ 1 keV and the wiggles at $\sim 2$--$3$   
keV (cp. with Fig.~\ref{gio1pl}).  }  
\label{bestfit_burst}   
\end{figure}

In comparison with the timing-mode observation (see Fig.~\ref{bestfit_burst}),    
the spectrum is steeper ($\Gamma\sim$2 instead   
of $\sim$1.7), and the reflection component much stronger 
($R\sim1.7$ instead of 0.3) and more   
ionized ($\xi\sim$ 3300 instead of 940 erg cm s$^{-1}$). The inner and outer radii   
are much smaller ($r_i <$20, $r_o \sim$100). The local absorber is slightly    
larger, while the ionized iron disk lines are not detected    
(upper limits to the EW of about 10 eV for   
both the He-- and H--like lines. The Fe H--like absorption line is still present,   
with a centroid energy of 6.98$\pm$0.02, i.e. consistent with both the   
rest frame value and the value    
obtained in the timing-mode observation. The EW is larger, about   
$-50$ eV, due to the fact that the line is now best fitted with a resolved width $\sigma$    
of 0.11$\pm$0.03 keV. The 1.85 and 2.24 keV emission lines   
are not required, with upper limits to the EWs of 5 and 2 eV, respectively.    
A 6.4 keV iron emission line is instead marginally found, with an EW   
of 6$\pm$4 eV.    
   
The above-mentioned results are consistent with a picture in which the disk is now more   
extended downwards and ionized   
(in fact, so ionized that no significant relativistic iron line   
is observed). The steeper spectrum can be due to the more efficient cooling, 
itself due to the   increase in soft disk photons. 
The fact that no thermal disk emission is required, however,   
implies that the values for the inner and outer radii must be used with caution. 
It must be noted that they are now determined 
only via the reflection component,    
with all the associated uncertainties   
(see discussion in the previous section), as no disk line is present.    
  
However, these results, especially those regarding  
the reflection components, can at least partly be an artifact due to calibration problems.  
A deficit of photons above 6$\div$7 keV, i.e. similar to what we find (see   
Fig.~\ref{bestfit_burst}), is indeed also present in the burst-mode observation of the   
Crab Nebula (Kirsch et al. 2005), a source where of course no significant   
reflection component is expected.
  
The 1 keV excess, which still persists, can again be well fitted by a power   
law plus an emission line, the latter    
with a flux slightly larger than that obtained in the timing-mode   
observation. In the reflection scenario discussed in the previous section, the optical depth   
is about a factor 10 larger, which implies a much lower EW of the line. These findings   
can be explained, at least qualitatively, by assuming that the matter is now thicker to the   
lines, with consequent self-absorption effects (see e.g. Matt et al. 1996).    
   
\begin{table}[t]   
\begin{center}   
\begin{tabular}{|c|c|}   
\hline   
& ~ \cr   
& {\bf Power law} \cr   
$\Gamma$ & 2.04$^{+0.01}_{-0.02}$ \cr   
& ~ \cr   
& {\bf Cold absorption} \cr   
$N_{\rm H,He,C,N,O}$ & 1.98$^{+0.02}_{-0.02}\times10^{22}$ cm$^{-2}$\cr   
& ~ \cr   
& {\bf Reflection} \cr   
$R/2\pi$ & 1.69$^{+0.16}_{-0.04}$ \cr   
$r_i/r_g$ & $<$20 \cr   
$r_o/r_g$ & 93$^{+11}_{-17}$ \cr   
$\xi$ & 3300$^{+600}_{-600}$ erg cm s$^{-1}$ \cr   
& ~ \cr   
& ~ \cr   
$F_{\rm 2-10~keV}$ &   
$\sim 6.6$ (unabs: $\sim 10.7$) $10^{-9}$ erg cm$^{-2}$ s$^{-1}$ \cr  
& ~ \cr  
$\chi^2$/d.o.f. & 248.2/219\cr   
& ~ \cr   
\hline   
\end{tabular}   
\caption{The best fit model parameters for the burst-mode observation (OBS2).    
See the text (Sect.~\ref{sec:burst}) for details. }   
\label{bestfittable_2}   
\end{center}   
\end{table}

\section{Conclusions}   
   
In this paper for the first time we have presented   
XMM-{\it Newton} observations of GRS\,1915+105.    
Several problems make it difficult   
to observe the source fruitfully with this satellite, first of all due   
to the extremely high flux -- in our case: $F_{\rm 2-10~keV} \sim 6.0\div6.6$    
(unabsorbed: $\sim 8.7\div10.7$) $\times 10^{-9}$ erg cm$^{-2}$ s$^{-1}$ --    
and to the notorious, erratic source variability,    
which does not help in triggering    
observations. We succedeed at both a) observing   
the source in a well-defined,  stable physical/spectral state and b)   
collecting EPIC-pn useful data, only marginally corrupted by telemetry problems.    
   
In both observations (April 17th and 21st, 2004),   
the source has been caught in a ``plateau" state, also characterized   
by stronger jet activity and higher radio level, which we identify    
with the conventional ``C'' state / $\chi$ variability mode as defined by    
Belloni \ea\ (1997a,b, 2000).    
In this state the source shows a QPO at $\sim 0.6$ Hz, i.e. what is expected    
in plateau intervals 
when the frequency vs. spectral hardness correlation is   
taken into account.   
   
In order to try to disentagle the different spectral components, and   
thus to better discriminate among the viable spectral/physical models,    
we used the $rms$ vs. $E$ method by Ponti \ea\ (2004).   
The resulting $rms$ is lower than 0.1 across the energy band.   
We further analyzed the variable $F_{\rm var}$, which takes the Poisson   
noise into account, and concluded that this variable is compatible with   
being null on the whole energy band, i.e. all spectral components are    
compatible with being constant during the two observations.   
  
A priori, some of the features in the spectrum may be affected by   
dust halo scattering, although this cannot explain the 1 keV 
excess (see Sect. 3.1.5). 
However, testing this hypothesis would first of   
all require an accurate analysis based on proper source imaging (if   
available); further, spectral modelling of such effects is not   
easy (see e.g. Costantini et al. 2005). Being beyond the scope   
of this paper, this study should be addressed in future works. 
   
We adopted a power law continuum model. This could be related to 
emission by a hot corona or to Comptonized thermal    
emission e.g. from the jet basis
(as proposed by Rodriguez \ea, 2004); however,    
an optically thick reflector is required to account for the smeared edge    
at $\sim 7$ keV (with a covering ratio of $\sim 0.4 \div 1.7 \times 2\pi$   
in the two observations, respectively). This may be evidence of an accretion    
disk being present, or just  
optically thick, only at quite a large distance from the central    
compact object, at least in the first observation    
($r_i/r_g > 300$ in OBS1, $\sim 20$ in OBS2).  
The relatively large amount of the reflection components implies that the   
primary X--ray emitting region should have a size that is at least   
comparable to the inner disk radius.  
That the disk is truncated, i.e. not present in the innermost part,   
is also suggested by the non-detection of a thermal disk emission component.    
    
Several line residuals are superimposed on the modeled continuum.   
Part of these may be due to calibration uncertainties, especially   
at the energies where changes in the EPIC effective area take place (e.g.   
1--3 keV). However, we found clear evidence of ionized iron emission   
around $\sim 7$ keV: data are well fitted with two ionized Fe K$\alpha$    
lines, possibly affected by mild relativistic broadening (being    
produced far away from the BH event horizon), plus a narrow absorption    
feature at $\sim 6.95$ keV.   
  
The puzzling presence of an intense, broad excess around   
1 keV can be explained in terms of reflection by an optically thin wind,
provided that the feature is real and that the discrepancies 
between the RGS and pn spectra (discussed in Sections 2.2.1 and 3.1.5) 
have to be attributed to RGS calibration problems. This assumption
is still unsubstantied, and should be regarded as a mere working hypothesis.   
The excess is indeed well fitted with a power law plus a line, unobscured   
by material that is intrinsic to the system. The centroid energy of the Gaussian    
line ($\sim$0.97 keV), its width (90 eV), and its EW against the reflected    
continuum (5.6 keV), together point to a blend of Ne K and Fe L lines.   
The value of the equivalent H column density is interestingly   
similar to the value of the obscuration by low Z elements (H, He, C, N, O) at    
the source core -- $N_{\rm H} \sim 1.6\times10^{22}$ cm$^{-2}$:   
in the disk wind hypothesis, this may   
therefore be taken as an upper limit to the {\it interstellar}   
matter column density. This value matches the expected galactic absorption 
in that direction well (Dickey \& Lockman 1990).   
   
On the other hand, a significant fraction of the absorber must   
be {\it local} to the source.   
We adopted a variable absorption model ({\sc varabs} in {\sc XSPEC}),    
assuming neutral matter    
and grouping the elements on the basis of both physical and practical   
considerations: elements which have probably a common origin, but also    
elements which are not very abundant (and therefore cannot be easily    
measured independently one from the other) with very abundant ones (e.g.    
Co and Ni with Fe): see the results in Table~\ref{bestfittable}.   
A significant overabundance of the heavier elements compared to the    
lighter ones is apparent, which suggests that a significant fraction of the    
absorber, traced by heavier species, is local to the source.    
These results are somewhat different from those derived from    
{\it Chandra} HETG data (Lee et al. 2002):    
in particular, Fe and S abundances are roughly the same, while   
our values for Mg are about twice, and 2/3 for Si.   
The overall scenario is, however, very similar.  
  
Clearly, the intrinsic absorption may be subject to    
substantial changes on longer timescales, as already observed with {\it Rossi}-XTE   
in similar plateaux (class $\chi$, state C):    
from a mean value of $\sim 4.5 \times10^{22}$ cm$^{-2}$,   
the total $N_{\rm H}$ may occasionally rise up to $\sim6 \times10^{22}$    
cm$^{-2}$ (Belloni \ea\ 2000).   
  
Assuming the customary value of $\sim 12.5$ kpc for the source's distance   
(Rodriguez \& Mirabel, 1999) we obtain an estimate of the intrinsic luminosity    
$L^{\rm unabs}_{\rm 2-10~keV} = 1.7\div2.0\times10^{38}$ erg s$^{-1}$ in the two   
observations.\footnote{Some authors claim that the distance could be   
substantially smaller    
(by up to a factor $\sim 1/2$: see Chapuis \& Corbel 2004), in which case    
a corresponding correction of the luminosity (and BH mass) estimate should    
be considered. }

\section{Acknowledgments}   
The authors would like to thank Stefano Bianchi, Laurence Boirin,
Massimo Cappi, Michal Dov\v{c}iak, Ken Ebisawa, Ya\"el Fuchs,   
Matteo Guainazzi, Mariano Mendez, Markus Kirsch, Jon Miller,    
Christian Motch, the anonymous referee,
and the whole XMM-{\it Newton} Helpdesk staff at Vilspa for    
many useful suggestions and technical help.    
Radio data at 15 GHz were observed with the Ryle Telescope, Cambridge,   
and kindly supplied by Guy Pooley.    
{\it Rossi}-XTE ASM (All Sky Monitor) results were provided by the teams    
at MIT and at the RXTE SOF and GOF at NASA's GSFC.   
AM wishes to thank the French Space Agency CNES, the University Paul
Sabatier in Toulouse, and the CNRS Research Group ``Ph\'enom\`enes Cosmiques    
de Haute Energie'' for financial support, as well as the Astronomical    
Institute of the Czech Academy of Sciences for their very kind hospitality.   
TB acknowledges partial support by grants INAF-PRIN 2002 and    
MIUR-PRIN 2003027534\_004.   
VK acknowledges the grant ref. GAAV IAA\,300030510.   
GP thanks the European Commission under the Marie Curie Early Stage 
Research  Training program for support.

\end{document}